%% file: eusipco2018.tex
\documentclass[conference]{IEEEtran}

\IEEEoverridecommandlockouts{}

\usepackage{cite}
\usepackage{amsmath,amssymb,amsfonts}
\usepackage{graphicx}
\usepackage{textcomp}
\usepackage{microtype}

\usepackage{booktabs}
\usepackage{nicefrac}
\usepackage[caption=false]{subfig}
\usepackage[inline]{enumitem}
\usepackage[group-four-digits]{siunitx}
\usepackage{url}
\usepackage{fancyhdr}

\usepackage{tikz}
\usetikzlibrary{arrows,automata,shapes}

\fancypagestyle{firststyle}
{
   \fancyhead[C]{\footnotesize 
           First published in the Proceedings of the 26th European Signal
           Processing Conference (EUSIPCO-2018) in 2018, published by EURASIP.
   }
}

\newcommand{\y}{y}
\newcommand{\ybar}{\bar{y}}
\newcommand{\Yset}{\mathcal{Y}}
\newcommand{\x}{\mathbf{x}}
\newcommand{\s}{s}
\newcommand{\compress}[1]{\mathcal{C}\left(#1\right)}
\DeclareMathOperator*{\argmax}{\arg\!\max}
\newcommand{\seq}[3]{#1_{#2:#3}}
\newcommand{\Y}{\seq{\y}{1}{T}}
\newcommand{\X}{\seq{\x}{1}{T}}

\title{Automatic Chord Recognition with Higher-Order Harmonic Language Modelling
\thanks{This work is supported by the European Research Council (ERC) under the EU's Horizon 2020 Framework Programme (ERC Grant Agreement number 670035, project ``Con Espressione'').}}

\author{\IEEEauthorblockN{Filip Korzeniowski and Gerhard Widmer}
\IEEEauthorblockA{Institute of Computational Perception,\\
Johannes Kepler University, Linz, Austria\\
Email: filip.korzeniowski@jku.at}
}

\begin{document}

\maketitle

\begin{abstract}
Common temporal models for automatic chord recognition model chord changes on a
frame-wise basis. Due to this fact, they are unable to capture musical
knowledge about chord progressions. In this paper, we propose a temporal model
that enables explicit modelling of chord changes and durations. We then apply
$N$-gram models and a neural-network-based acoustic model within this
framework, and evaluate the effect of model overconfidence. Our results show
that model overconfidence plays only a minor role (but target smoothing
still improves the acoustic model), and that stronger chord language models do
improve recognition results, however their effects are small compared to other
domains.
\end{abstract}

\begin{IEEEkeywords}
Chord Recognition, Language Modelling, N-Grams, Neural Networks
\end{IEEEkeywords}

Research on automatic chord recognition has recently focused on improving
frame-wise predictions of acoustic
models~\cite{korzeniowski_feature_2016,mcfee_structured_2017,humphrey_learning_2012}.
This trend roots in the fact that existing temporal models just smooth the
predictions of an acoustic model, and do not incorporate musical
knowledge~\cite{cho_relative_2014}. As we argue
in~\cite{korzeniowski_futility_2017}, the reason is that such temporal models
are usually applied on the audio-frame level, where even non-Markovian models
fail to capture musical properties.  

We know the importance of language models in domains such as speech
recognition, where hierarchical grammar, pronunciation and context models
reduce word error rates by a large margin. However, the degree to which
higher-order language models improve chord recognition results yet remains
unexplored. In this paper, we want to shed light on this question. Motivated by
the preliminary results from~\cite{korzeniowski_futility_2017}, we show how to
integrate chord-level harmonic language models into a chord recognition system,
and evaluate its properties.

Our contributions in this paper are as follows. We present a probabilistic
model that allows for combining an acoustic model with explicit modelling of
chord transitions and chord durations. This allows us to deploy language models
on the \emph{chord level}, not the frame level. Within this framework, we then
apply $N$-gram chord language models on top of an neural network based acoustic
model. Finally, we evaluate to which degree this combination suffers from
acoustic model over-confidence, a typical problem with neural acoustic
models~\cite{chorowski_better_2016}.

\thispagestyle{firststyle}
\section{Problem Definition}

Chord recognition is a sequence labelling problem similar to speech
recognition. In contrast to the latter, we are also interested in the start
and end points of the segments. Formally, assume $\seq{\x}{1}{T}$\footnote{We use the notation $\seq{\mathbf{v}}{i}{j}$ to indicate $(\mathbf{v}_i, \mathbf{v}_{i+1}, \ldots, \mathbf{v}_j)$.}
is a time-frequency representation of the input signal; the goal is then to
find $\seq{\y}{1}{T}$, where $\y_t \in \Yset$ is a chord symbol
from a chord vocabulary $\Yset$, such that $\y_t$ is the correct
harmonic interpretation of the audio content represented by $\x_t$.
Formulated probabilistically, we want to infer
\begin{align}
        \seq{\hat{\y}}{1}{T} &= \argmax_{\seq{\y}{1}{T}} P(\seq{\y}{1}{T} \mid \seq{\x}{1}{T}).
        \label{eq:decoding}
\end{align}

Assuming a generative structure where $\Y$ is a left-to-right process, and each $\x_t$ only depends on $\y_t$,
\begin{align*}
        P(\Y \mid \X) &\propto \prod_t \frac{1}{P(\y_t)} P_A\left(\y_t \mid \x_t\right) P_T\left(\y_t \mid \seq{\y}{1}{t-1}\right),
\end{align*}
where the $\nicefrac{1}{P(\y_t)}$ is a label prior that we assume uniform for
simplicity~\cite{renals_connectionist_1994}, $P_A(\y_t \mid \x_t)$ is the
\emph{acoustic model}, and $P_T\left(\y_t \mid \seq{\y}{1}{t-1}\right)$ the \emph{temporal
model}.

Common choices for $P_T$ (e.g.\ Markov processes or recurrent neural networks)
are unable to model the underlying musical language of harmony meaningfully. As
shown in~\cite{korzeniowski_futility_2017}, this is because modelling the
symbolic chord sequence on a frame-wise basis is dominated by self-transitions.
This prevents the models from learning higher-level knowledge about chord
changes. To avoid this, we disentangle $P_T$ into a chord
\emph{language model} $P_L$, and a chord \emph{duration model} $P_D$.

The chord language model is defined as $P_L\left(\ybar_i \mid \seq{\ybar}{1}{i-1}\right)$,
where $\seq{\ybar}{1}{i} = \compress{\y_{1:t}}$, and $\compress{\cdot}$ is a sequence compression mapping that removes all
consecutive duplicates of a symbol (e.g.\, $\compress{(a,a,b,b,a)} = (a,b,a)$).
$P_L$ thus only considers chord \emph{changes}.
The duration model is defined as $P_D\left(\s_t \mid \y_{1:t-1}\right)$, where
$\s_t \in \{\text{s}, \text{c}\}$ indicates whether the chord changes (c) or
stays the same (s) at time $t$. $P_D$ thus only considers chord \emph{durations}.
The temporal model is then formulated as:
\newcommand{\past}{\seq{\y}{1}{t-1}}
\newcommand{\pastlang}{\seq{\ybar}{1}{i-1}}
\begin{align}
        P_T&\left(\y_t \mid \past\right) =  \label{eq:temporal_model} \\
                                         &\begin{cases}
          P_L\left(\ybar_i \mid \pastlang\right) P_D\left(\text{c} \mid \past \right) & \text{if } y_t \neq y_{t-1} \\
    P_D\left(\text{s} \mid \past \right) & \text{else} \nonumber
  \end{cases}.
\end{align}

To fully specify the system, we need to define the acoustic model $P_A$,
the language model $P_L$, and the duration model $P_D$.

\section{Models}

\subsection{Acoustic Model}

The acoustic model used in this paper is a minor variation of the one
introduced in~\cite{korzeniowski_fully_2016}. It is a
VGG-style~\cite{simonyan_very_2014} fully convolutional neural network with 3
convolutional blocks: the first consists of 4 layers of 32 3$\times$3 filters,
followed by $2\times1$ max-pooling in frequency; the second comprises 2 layers
of 64 such filters followed by the same pooling scheme; the third is a single
layer of 128 12$\times$9 filters.  Each of the blocks is followed by
feature-map-wise dropout with probability 0.2, and each layer is followed by
batch normalisation~\cite{ioffe_batch_2015} and an exponential linear
activation function~\cite{clevert_fast_2016}. Finally, a linear convolution
with 25 1$\times$1 filters followed by global average pooling and a softmax
produces the chord class probabilities $P_A(\y_k \mid \x_k)$.  The input to the
network is a log-magnitude log-frequency spectrogram patch of 1.5 seconds.
See~\cite{korzeniowski_fully_2016} for a detailed description of the input
processing and training schemes.

Neural networks tend to produce overconfident predictions, which leads to
probability distributions with high peaks. This causes a weaker training
signal because the loss function saturates, and makes the acoustic model
dominate the language model at test time~\cite{chorowski_better_2016}. Here,
we investigate two approaches to mitigate these effects: using a temperature
softmax in the classification layer of the network, and training using smoothed
labels.

The temperature softmax replaces the regular softmax activation function at
test time with
\begin{align*}
    {\sigma\left(\mathbf{z}\right)}_j = \frac{e^{\nicefrac{z_j}{T}}}{\sum_{k=1}^K e^{\nicefrac{z_k}{T}}},
\end{align*}
where $\mathbf{z}$ is a real vector. High values for $T$ make the resulting
distribution smoother. With $T=1$, the function corresponds to the standard
softmax. The advantage of this method is that the network does not need to be
retrained.

Target smoothing, on the other hand, trains the network with with a smoothed
version of the target labels. In this paper, we explore three ways of
smoothing: \emph{uniform smoothing}, where a proportion of $1 - \beta$ of the
correct probability is assigned uniformly to the other classes; \emph{unigram
smoothing}, where the smoothed probability is assigned according to the class
distribution in the training set~\cite{szegedy_rethinking_2015}; and
\emph{target smearing}, where the target is smeared in time using a running
mean filter. The latter is inspired by a similar approach
in~\cite{ullrich_boundary_2014} to counteract inaccurate segment boundary
annotations.


\subsection{Language Model}

We designed the temporal model in Eq.~\ref{eq:temporal_model} in a way that
enables chord changes to be modelled explicitly via $P_L(\ybar_k \mid
\compress{\seq{\ybar}{1}{k-1}})$. This formulation allows to use all past
chords to predict the next. While this is a powerful and general notion, it
prohibits efficient exact decoding of the sequence. We would have to rely on
approximate methods to find $\seq{\hat{\y}}{1}{T}$ (Eq.~\ref{eq:decoding}). However, we can restrict
the number of past chords the language model can consider, and use
higher-order Markov models for exact decoding. To achieve that, we use
$N$-grams for language modelling in this work.

$N$-gram language models are Markovian probabilistic models that assume only a
fixed-length history (of length $N - 1$) to be relevant for predicting the next
symbol.  This fixed-length history allows the probabilities to be stored in a
table, with its entries computed using maximum-likelihood estimation
(MLE)---i.e., by counting occurrences in the training set.

With larger $N$, the sparsity of the probability table increases exponentially,
because we only have a finite number of $N$-grams in our training set. We
tackle this problem using Lidstone smoothing, and add a pseudo-count $\alpha$
to each possible $N$-gram. We determine the best value for $\alpha$ for each
model using the validation set.

\begin{figure}
\centering
\input{figs/negbinhmm.tex}
\caption{Markov chain modelling the duration of a chord segment ($K=3$). 
        The probability of staying in one of the states follows 
        a negative binomial distribution.}
\label{fig:negbinhmm}
\end{figure}
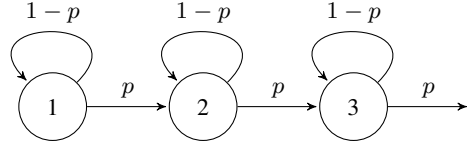

\subsection{Duration Model}

\begin{figure}[t!]
\centering
\includegraphics{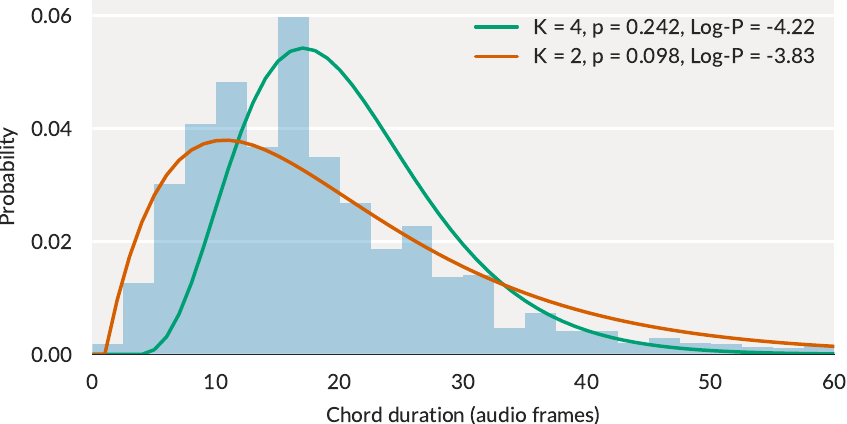}
\caption{Histogram of chord durations with two configurations
of the negative binomial distribution. The log-probability is computed on a
validation fold.}
\label{fig:duration_pmf}
\end{figure}

The focus of this paper is on how to meaningfully incorporate chord language
models beyond simple first-order transitions. We thus use only a simple
duration model based on the negative binomial distribution, with the probability
mass function
\begin{align*}
P(k) = \binom{k+K-1}{K-1} p^K (1 - p)^k,
\end{align*}
where $K$ is the number of failures, $p$ the failure probability, and $k$ the
number of successes given $K$ failures. For our purposes, $k + K$ is the length
of a chord in audio frames.

The main advantage
of this choice is that a negative binomial distribution is easily represented
using only few states in a HMM (see Fig.~\ref{fig:negbinhmm}), while still
reasonably modelling the length of chord segments (see
Fig.~\ref{fig:duration_pmf}). For simplicity, we use the same duration model
for all chords. The parameters ($K$, the number of states used for modelling
the duration, and $p$, the probability of moving to the next state) are
estimated using MLE.\@

\subsection{Model Integration}

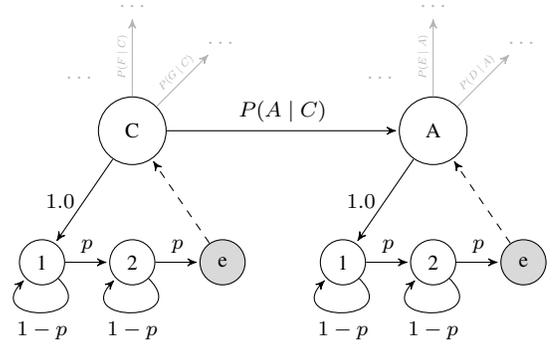
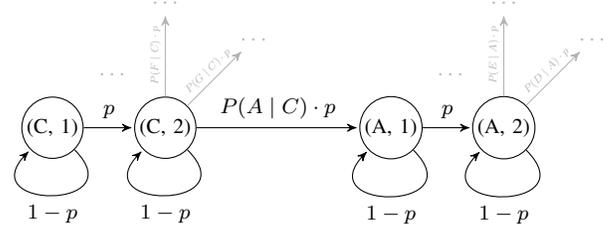
\begin{figure}[t!]
    \centering
    \subfloat[First-Order Hierarchical HMM.]{\label{fig:hohhmm}
        \input{figs/hohmm.tex}
    }\\
    \subfloat[Flattened version of the First-Order Hierarchical HMM.]{\label{fig:flattened_hohhmm}
        \input{figs/flattened_hohhmm.tex}
    }
    \caption{Exemplary Hierarchical HMM and its flattened version. We left out
            incoming and outgoing transitions of the chord states for clarity
    (except $\text{C}\rightarrow\text{A}$ and the ones indicated in gray). The
    model uses 2 states for duration modelling, with ``e'' referring to the
    final state on the duration level (see~\cite{fine_hierarchical_1998} for 
    details). Although we depict a first-order
    language model here, the same transformation works for higher-order models.}
\end{figure}

If we combine an $N$-gram language model with a negative binomial duration
model, the temporal model $P_T$ becomes a Hierarchical Hidden Markov
Model~\cite{fine_hierarchical_1998} with a higher-order Markov model on the top
level (the language model) and a first-order HMM at the second level (see
Fig.~\ref{fig:hohhmm}). We can translate the hierarchical HMM into a
first-order HMM\@; this will allow us to use many existing and optimised HMM
implementations.

To this end, we first transform the higher-order HMM on the top level into a
first-order one as shown e.g.\ in~\cite{hadar_highorder_2009}: we factor
the dependencies beyond first-order into the HMM state, considering that
self-transitions are impossible as
\begin{align*}
\Yset_N = \left\{(y_1, \ldots, y_N): y_i \in \Yset, y_i \neq y_{i+1} \right\},
\end{align*}
where $N$ is the order of the $N$-gram model. Semantically, $(y_1, \ldots, y_{N})$
represents chord $y_1$, having seen $y_2, \ldots, y_{N}$ in the immediate
past. This increases the number of states from $|\Yset|$ to
$|\Yset| \cdot {(|\Yset| - 1)}^{N-1}$.

We then flatten out the
hierarchical HMM by combining the state spaces of both levels as
$\Yset_N \times [1..K]$, and connecting all incoming transitions of a
chord state to the corresponding first duration state, and all outgoing
transitions from the last duration state (where the outgoing probabilities
are multiplied by $p$). Formally,
\begin{align*}
        \Yset_N^{(K)} &= \left\{(\mathbf{y}, k): \mathbf{y} \in \Yset_N, k \in [1..K]\right\},
\end{align*}
with the transition probabilities defined as
\newcommand{\yp}{y^\prime}
\begin{align*}
        P((\mathbf{y}, k) \mid (\mathbf{y}, k)) &= 1 - p, \\
        P((\mathbf{y}, k + 1) \mid (\mathbf{y}, k)) &= p, \\
        P((\mathbf{y}, 1) \mid (\mathbf{\yp}, K)) &= P_L(y_1 \mid y_{2:N}) \cdot p,
\end{align*}
where $y_{2:N} = \yp_{1:N-1}$. All other transitions have zero probability.
Fig.~\ref{fig:flattened_hohhmm} shows the HMM from Fig.~\ref{fig:hohhmm} after
the transformation.

\begin{figure*}[ht!]
\vspace{0.15in}
\centering
\includegraphics{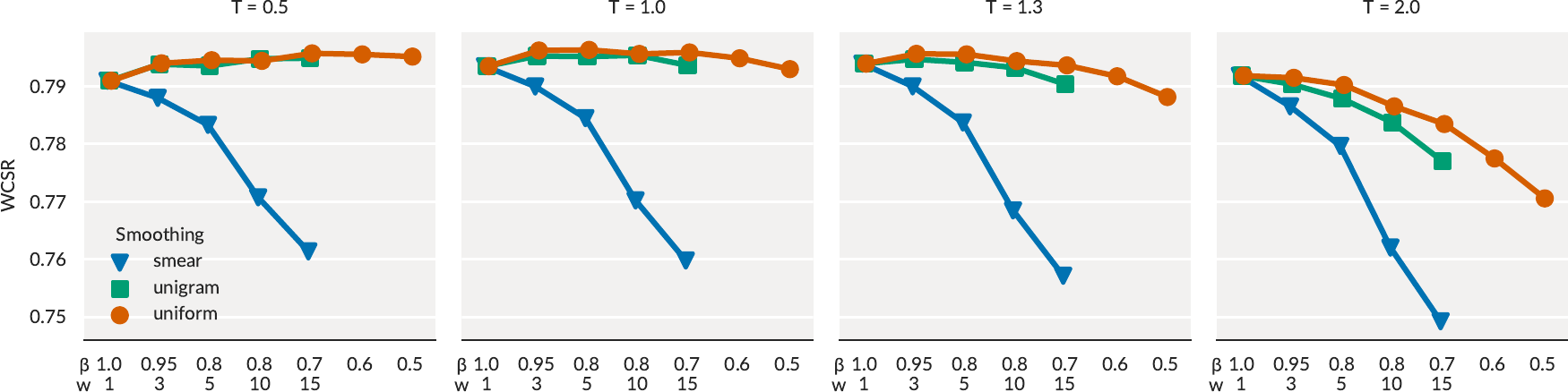}
\captionsetup{singlelinecheck=off}
\caption[dummy stuff]{%
The effect of temperature $T$, smoothing type, and smoothing intensity on the WCSR\@.
The x-axis shows the smoothing intensity: for uniform and unigram smoothing,
$\beta$ indicates how much probability mass was kept at the true label during
training; for target smearing, $w$ is the width of the running mean filter
used for smearing the targets in time. For these results, a 2-gram language model was used,
but the outcomes are similar for other language models. The key observations
are the following:
\begin{enumerate*}[label= (\roman*)]
\item target smearing is always detrimental;
\item uniform smoothing works slightly better than unigram smoothing (in other domains, authors report the contrary~\cite{chorowski_better_2016}); and
\item smoothing improves the results, however, excessive smoothing is harmful in combination with higher softmax temperatures (a relation we explore in greater detail in Fig.~\ref{fig:smoothing_temp_lm}).
\end{enumerate*}}\label{fig:smoothing_types}
\end{figure*}

\begin{figure*}[ht!]
\includegraphics[width=\textwidth]{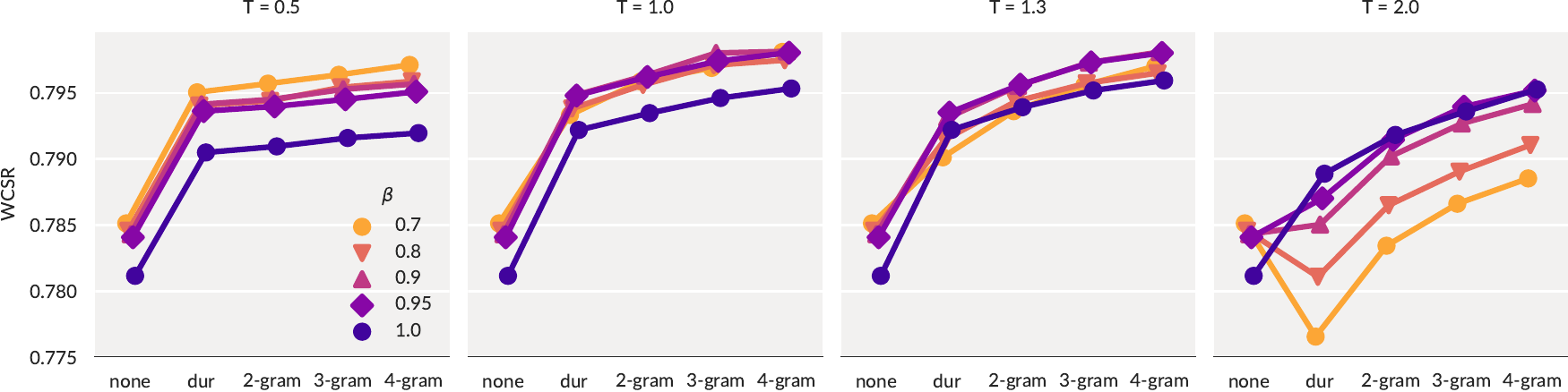}
\caption[dummy stuff]{%
Interaction of temperature $T$, smoothing intensity $\beta$ and language model
with respect to the WCSR\@. We show four language model configurations:
\emph{none} means using the predictions of the acoustic model directly;
\emph{dur} means using the chord duration model, but no chord language model;
and \emph{$N$-gram} means using the duration model with the respective language
model. Here, we only show results using uniform smoothing, which turned out to
be the best smoothing technique we examined in this paper (see Fig.~\ref{fig:smoothing_types}).
We observe the following:
\begin{enumerate*}[label= (\roman*)]
        \item Even simple duration modelling accounts for the majority of the
              improvement (in accordance with~\cite{chen_chord_2012}).
        \item Chord language models further improve the results---the stronger
              the language model, the bigger the improvement.
        \item Temperature and smoothing interact: at $T=1$, the amount of
              smoothing plays only a minor role; if we lower $T$ (and thus make
              the predictions more confident), we need stronger smoothing to
              compensate for that; if we increase both $T$ and the smoothing
              intensity, the predictions of the acoustic model are over-ruled
              by the language model, which shows to be detrimental.
        \item Smoothing has an additional effect during the training of the
              acoustic model that cannot be achieved using post-hoc changes
              in softmax temperature. Unsmoothed models never achieve the best
              result, regardless of $T$.
\end{enumerate*}
}\label{fig:smoothing_temp_lm}
\end{figure*}

The resulting model is similar to a higher-order duration-explicit HMM (DHMM).
The main difference is that we use a compact duration model that can
assign duration probabilities using few states, while standard DHMMs do not
scale well if longer durations need to be modelled (their computation
increases by a factor of $\nicefrac{D^2}{2}$, where $D$ is the longest
duration to be modelled~\cite{rabiner_tutorial_1989}). For example, 
\cite{chen_chord_2012} uses first-order DHMMs to decode beat-synchronised
chord sequences, with $D=20$. In our case, we would need a much higher $D$,
since our model operates on the frame level, which would result in a
prohibitively large state space. In comparison, our duration models use only
$K=2$ (as determined by MLE) states to model the duration, which significantly reduces the
computational burden.

\section{Experiments}

Our experiments aim at uncovering
\begin{enumerate*}[label= (\roman*)]
\item if acoustic model overconfidence is a problem in this scenario,
\item whether smoothing techniques can mitigate it, and
\item whether and to which degree chord language modelling improves chord
      recognition results.
\end{enumerate*}
To this end, we investigated the effect of various parameters: softmax
temperature $T \in \{0.5, 1.0, 1.3, 2.0\}$, smoothing type (uniform,
unigram, and smear), smoothing intensity $\beta \in \{0.5, 0.6, 0.7,
\allowbreak 0.8, 0.9, 0.95\}$ and smearing width $w \in \{3, 5, 10, 15\}$, and
the language model order $N \in \{2, 3, 4\}$.

The experiments were carried out using 4-fold cross-validation on a compound
dataset consisting of the following sub-sets:
\textbf{Isophonics\footnote{\url{http://isophonics.net/datasets}}:} 180 songs
by the Beatles, 19 songs by Queen, and 18 songs by Zweieck, 10:21
hours of audio; \textbf{RWC Popular~\cite{goto_rwc_2002}:} 100 songs in the style of
American and Japanese pop music, 6:46 hours of audio; \textbf{Robbie
Williams~\cite{digiorgi_automatic_2013}:} 65 songs by Robbie Williams, 4:30 of audio; and
\textbf{McGill Billboard~\cite{burgoyne_expert_2011}:} 742 songs sampled from the American
billboard charts between 1958 and 1991, 44:42 hours of audio. The compound
dataset thus comprises 1125 unique songs, and a total of 66:21 hours of audio.

We focus on the major/minor chord vocabulary (i.e.\ major and minor chords for
each of the 12 semitones, plus a ``no-chord'' class, totalling 25 classes). 
The evaluation measure we are interested in is thus the weighted chord symbol
recall of major and minor chords, $\text{WCSR} = \nicefrac{t_c}{t_a}$,
where $t_c$ is the total time the our system recognises the correct chord, and
$t_a$ is the total duration of annotations of the chord types of interest.

\subsection{Results and Discussion}

We analyse the interactions between temperature, smoothing, and language
modelling in Fig.~\ref{fig:smoothing_types} and
Fig.~\ref{fig:smoothing_temp_lm}. Uniform smoothing seems to perform best,
while increasing the temperature in the softmax is unnecessary if smoothing
is used. On the other hand, target smearing performs poorly; it is thus not
a proper way to cope with uncertainty in the annotated chord boundaries.

The results indicate that in our scenario, acoustic model overconfidence is not
a major issue. The reason might be that the temporal model we use in this work
allows for exact decoding. If we were forced to perform approximate inference
(e.g.\ by using a RNN-based language model), this overconfidence could cut off
promising paths early. Target smoothing still exhibits a positive effect during
the training of the acoustic model, and can be used to fine-balance the
interaction between acoustic and temporal models.

Further, we see consistent improvement the stronger the language model is
(i.e., the higher $N$ is). Although we were not able to evaluate models beyond
$N=4$ for all configurations, we ran a 5-gram model on the best configuration
for $N=4$. The results are shown in Table~\ref{tab:results}. 

\begin{table}[t]
\sisetup{round-mode=places,round-precision=2}
\centering
\caption{WCSR for the compound dataset. For these results, we use a softmax temperature of $T=1.0$ and
uniform smoothing with $\beta=0.9$.\vspace{-0.4em}}\label{tab:results}
\begin{tabular}{cccccc}
        \toprule
        \textbf{None} & \textbf{Dur.} & \textbf{2-gram} & \textbf{3-gram} & \textbf{4-gram} & \textbf{5-gram} \\
        \midrule
        \num{78.5149} & \num{79.3339} & \num{79.5870} & \num{79.6895} & \num{79.8105} & \num{79.8753}\\
        \bottomrule
\end{tabular}
\end{table}

Although consistent, the improvement is marginal
compared to the effect language models show in other domains such as
speech recognition. There are two possible interpretations of this result:
\begin{enumerate*}[label= (\roman*)]
\item even if modelled explicitly, chord language models contribute little to
      the final results, and the most important part is indeed modelling the
      chord duration; and
\item the language models used in this paper are simply not good enough to
      make a major difference.
\end{enumerate*}
While the true reason yet remains unclear, the structure of the temporal model
we propose enables us to research both possibilities in future work, because it
makes their contributions explicit.

Finally, our results confirm the importance of duration
modelling~\cite{chen_chord_2012}. Although the duration model we use here is
simplistic, it improves results considerably. However, in further informal
experiments, we found that it underestimates the probability of long chord
segments, which impairs results. This indicates that there is still potential
for improvement in this part of our model.

\section{Conclusion}

We proposed a probabilistic structure for the temporal model of chord
recognition systems. This structure disentangles a chord language model from a
chord duration model. We then applied $N$-gram chord language models within
this structure and evaluated various properties of the system. The key outcomes
are that 
\begin{enumerate*}[label= (\roman*)]
\item acoustic model overconfidence plays only a minor role (but target
        smoothing still improves the acoustic model),
\item chord duration modelling (or, sequence smoothing) improves results considerably,
        which confirms prior studies~\cite{chen_chord_2012,cho_relative_2014}, and
\item while employing $N$-gram models also improves the results, their effect
      is marginal compared to other domains such as speech recognition.
\end{enumerate*}

Why is this the case? Static $N$-gram models might only capture global
statistics of chord progressions, and these could be too general to guide and
correct predictions of the acoustic model. More powerful models may be required.
As shown in~\cite{korzeniowski_largescale_2018}, RNN-based chord language
models are able to adapt to the currently processed song, and thus might be
more suited for the task at hand.

The proposed probabilistic structure thus opens various possibilities for
future work. We could explore better language models, e.g.\ by using more
sophisticated smoothing techniques, RNN-based models, or probabilistic models
that take into account the key of a song (the probability of chord transitions 
varies depending on the key).
More intelligent duration models could take into account the tempo and harmonic 
rhythm of a song (the rhythm in which chords change). Using the model
presented in this paper, we could then link the improvements of each
individual model to improvements in the final chord recognition score.

\bibliographystyle{IEEEtran}
\bibliography{eusipco2018}

\end{document}

%% file: figs/negbinhmm.tex
\begin{tikzpicture}[->,>=stealth',shorten >=1pt,auto]
\tikzset{font=\small}
\tikzstyle{every state}=[fill=none,draw=black,text=black,minimum size=0.9cm,inner sep=1pt, node distance=2cm]

\node[state]                       (h0) {1};
\node[state,right of=h0]  (h1) {2};
\node[state,right of=h1]  (h2) {3};
\node[state,right of=h2,draw=none] (ht) {};

\path (h0) edge node {$p$} (h1) edge [loop above,out=45,in=135,looseness=5] node {$1 - p$} (h0)
	    (h1) edge node {$p$} (h2) edge [loop above,out=45,in=135,looseness=5] node {$1 - p$} (h1)
			(h2) edge node {$p$} (ht) edge [loop above,out=45,in=135,looseness=5] node {$1 - p$} (h2);

\end{tikzpicture}

%% file: figs/hohmm.tex
\begin{tikzpicture}[->,>=stealth',shorten >=1pt,auto]
\tikzset{font=\footnotesize}
\tikzstyle{every state}=[fill=none,draw=black,text=black,minimum size=0.6cm, node distance=1.2cm]

\node[state,minimum size=0.9cm]              (C) {C};
\node[state,below of=C, node distance=1.75cm] (C2) {2};
\node[state,left of=C2]                      (C1) {1};
\node[state,right of=C2,fill=black!15]       (Ce) {e};
\path (C1) edge node {$p$} (C2) edge [loop below, out=315, in=225,looseness=5] node {$1 - p$} (C1)
      (C2) edge node {$p$} (Ce) edge [loop below, out=315, in=225,looseness=5] node {$1 - p$} (C2)
      (C) edge [left] node {$1.0$} (C1)
      (Ce) edge [dashed] (C);

\node[state,minimum size=0.9cm,right of=C, node distance=4cm]              (A) {A};
\node[state,below of=A, node distance=1.75cm] (A2) {2};
\node[state,left of=A2]                      (A1) {1};
\node[state,right of=A2,fill=black!15]       (Ae) {e};
\path (A1) edge node {$p$} (A2) edge [loop below, out=315, in=225,looseness=5] node {$1 - p$} (A1)
      (A2) edge node {$p$} (Ae) edge [loop below, out=315, in=225,looseness=5] node {$1 - p$} (A2)
      (A) edge [left] node {$1.0$} (A1)
      (Ae) edge [dashed] (A);

\node[above right of=C,text=black!30,node distance=1.66cm] (G) {$\ldots$};
\node[above of=C,text=black!30,node distance=1.66cm] (F) {$\ldots$};
\node[above left of=C,text=black!30,node distance=1cm] {$\ldots$};

\path (C) edge node {$P(A\mid C)$} (A)
          edge [color=black!30] node[scale=0.5,sloped, anchor=center, above, text=black!30,color=black!30]
                                {$P(G\mid C)$} (G)
          edge [color=black!30] node[scale=0.5,sloped, anchor=center, above, text=black!30,color=black!30]
                                {$P(F\mid C)$} (F);

\node[above right of=A,text=black!30,node distance=1.66cm] (D) {$\ldots$};
\node[above of=A,text=black!30,node distance=1.66cm] (E) {$\ldots$};
\node[above left of=A,text=black!30,node distance=1cm] {$\ldots$};

\path (A) edge [color=black!30] node[scale=0.5,sloped, anchor=center, above, text=black!30,color=black!30]
                                {$P(E\mid A)$} (E)
          edge [color=black!30] node[scale=0.5,sloped, anchor=center, above, text=black!30,color=black!30]
                                {$P(D\mid A)$} (D);

\end{tikzpicture}

%% file: figs/flattened_hohhmm.tex
\begin{tikzpicture}[->,>=stealth',shorten >=1pt,auto]
\tikzset{font=\footnotesize}
\tikzstyle{every state}=[fill=none,draw=black,text=black,minimum size=0.6cm,
                         inner sep=1pt, node distance=1.5cm]

\node[state]            (c1) {(C, 1)};
\node[state,right of=c1]   (c2) {(C, 2)};
\node[state,right of=c2,node distance=3cm]   (a1) {(A, 1)};
\node[state,right of=a1]   (a2) {(A, 2)};

\path (c1) edge node {$p$} (c2) edge [loop below, out=315, in=225,looseness=5] node {$1 - p$} (c1)
      (c2) edge node {$P(A\mid C) \cdot p$} (a1) edge [loop below, out=315, in=225,looseness=5] node {$1 - p$} (c2)
      (a1) edge node {$p$} (a2) edge [loop below, out=315, in=225,looseness=5] node {$1 - p$} (a1)
      (a2) edge [loop below, out=315, in=225, looseness=5] node {$1 - p$} (a2);

\node[above right of=c2,text=black!30,node distance=1.66cm] (g1) {$\ldots$};
\node[above of=c2,text=black!30,node distance=1.66cm] (f1) {$\ldots$};
\node[above left of=c2,text=black!30,node distance=1cm] {$\ldots$};

\path (c2) edge [color=black!30] node[scale=0.5,sloped, anchor=center, above, text=black!30,color=black!30]
                                {$P(G\mid C)\cdot p$} (g1)
      (c2) edge [color=black!30] node[scale=0.5,sloped, anchor=center, above, text=black!30,color=black!30]
                                {$P(F\mid C)\cdot p$} (f1);

\node[above right of=a2,text=black!30,node distance=1.66cm] (d1) {$\ldots$};
\node[above of=a2,text=black!30,node distance=1.66cm] (e1) {$\ldots$};
\node[above left of=a2,text=black!30,node distance=1cm] {$\ldots$};

\path (a2) edge [color=black!30] node[scale=0.5,sloped, anchor=center, above, text=black!30,color=black!30]
                                {$P(D\mid A)\cdot p$} (d1)
      (a2) edge [color=black!30] node[scale=0.5,sloped, anchor=center, above, text=black!30,color=black!30]
                                {$P(E\mid A)\cdot p$} (e1);

\end{tikzpicture}

%% file: eusipco2018.bbl
\begin{thebibliography}{10}
\providecommand{\url}[1]{#1}
\csname url@samestyle\endcsname
\providecommand{\newblock}{\relax}
\providecommand{\bibinfo}[2]{#2}
\providecommand{\BIBentrySTDinterwordspacing}{\spaceskip=0pt\relax}
\providecommand{\BIBentryALTinterwordstretchfactor}{4}
\providecommand{\BIBentryALTinterwordspacing}{\spaceskip=\fontdimen2\font plus
\BIBentryALTinterwordstretchfactor\fontdimen3\font minus
  \fontdimen4\font\relax}
\providecommand{\BIBforeignlanguage}[2]{{%
\expandafter\ifx\csname l@#1\endcsname\relax
\typeout{** WARNING: IEEEtran.bst: No hyphenation pattern has been}%
\typeout{** loaded for the language `#1'. Using the pattern for}%
\typeout{** the default language instead.}%
\else
\language=\csname l@#1\endcsname
\fi
#2}}
\providecommand{\BIBdecl}{\relax}
\BIBdecl

\bibitem{korzeniowski_feature_2016}
F.~Korzeniowski and G.~Widmer, ``Feature {{Learning}} for {{Chord
  Recognition}}: {{The Deep Chroma Extractor}},'' in \emph{17th {{International
  Society}} for {{Music Information Retrieval Conference}} ({{ISMIR}})}, New
  York, USA, Aug. 2016.

\bibitem{mcfee_structured_2017}
B.~McFee and J.~P. Bello, ``Structured {{Training}} for {{Large}}-{{Vocabulary
  Chord Recognition}},'' in \emph{18th {{International Society}} for {{Music
  Information Retrieval Conference}} ({{ISMIR}})}, Suzhou, China, Oct. 2017.

\bibitem{humphrey_learning_2012}
E.~J. Humphrey, T.~Cho, and J.~P. Bello, ``Learning a {{Robust
  Tonnetz}}-{{Space Transform}} for {{Automatic Chord Recognition}},'' in
  \emph{2012 {{IEEE International Conference}} on {{Acoustics}}, {{Speech}} and
  {{Signal Processing}} ({{ICASSP}})}, Kyoto, Japan, 2012.

\bibitem{cho_relative_2014}
T.~Cho and J.~P. Bello, ``On the {{Relative Importance}} of {{Individual
  Components}} of {{Chord Recognition Systems}},'' \emph{IEEE/ACM Transactions
  on Audio, Speech, and Language Processing}, vol.~22, no.~2, pp. 477--492,
  Feb. 2014.

\bibitem{korzeniowski_futility_2017}
F.~Korzeniowski and G.~Widmer, ``On the {{Futility}} of {{Learning Complex
  Frame}}-{{Level Language Models}} for {{Chord Recognition}},'' in
  \emph{Proceedings of the {{AES International Conference}} on {{Semantic
  Audio}}}, Erlangen, Germany, Jun. 2017.

\bibitem{chorowski_better_2016}
J.~Chorowski and N.~Jaitly, ``Towards better decoding and language model
  integration in sequence to sequence models,'' \emph{arXiv:1612.02695}, Dec.
  2016.

\bibitem{renals_connectionist_1994}
S.~Renals, N.~Morgan, H.~Bourlard, M.~Cohen, and H.~Franco, ``Connectionist
  {{Probability Estimators}} in {{HMM Speech Recognition}},'' \emph{IEEE
  Transactions on Speech and Audio Processing}, vol.~2, no.~1, pp. 161--174,
  Jan. 1994.

\bibitem{korzeniowski_fully_2016}
F.~Korzeniowski and G.~Widmer, ``A {{Fully Convolutional Deep Auditory Model}}
  for {{Musical Chord Recognition}},'' in \emph{26th {{IEEE International
  Workshop}} on {{Machine Learning}} for {{Signal Processing}} ({{MLSP}})},
  Salerno, Italy, Sep. 2016.

\bibitem{simonyan_very_2014}
K.~Simonyan and A.~Zisserman, ``Very {{Deep Convolutional Networks}} for
  {{Large}}-{{Scale Image Recognition}},'' \emph{arXiv:1409.1556}, Sep. 2014.

\bibitem{ioffe_batch_2015}
S.~Ioffe and C.~Szegedy, ``Batch {{Normalization}}: {{Accelerating Deep Network
  Training}} by {{Reducing Internal Covariate Shift}},''
  \emph{arXiv:1502.03167}, Mar. 2015.

\bibitem{clevert_fast_2016}
D.-A. Clevert, T.~Unterthiner, and S.~Hochreiter, ``Fast and {{Accurate Deep
  Network Learning}} by {{Exponential Linear Units}} ({{ELUs}}),'' in
  \emph{International {{Conference}} on {{Learning Representations}}
  ({{ICLR}}), {{arXiv}}:1511.07289}, San Juan, Puerto Rico, Feb. 2016.

\bibitem{szegedy_rethinking_2015}
C.~Szegedy, V.~Vanhoucke, S.~Ioffe, J.~Shlens, and Z.~Wojna, ``Rethinking the
  {{Inception Architecture}} for {{Computer Vision}},''
  \emph{arXiv:1512.00567}, Dec. 2015.

\bibitem{ullrich_boundary_2014}
K.~Ullrich, J.~Schl{\"u}ter, and T.~Grill, ``Boundary {{Detection}} in {{Music
  Structure Analysis Using Convolutional Neural Networks}},'' in \emph{15th
  {{International Society}} for {{Music Information Retrieval Conference}}
  ({{ISMIR}})}, Taipei, Taiwan, Oct. 2014.

\bibitem{fine_hierarchical_1998}
S.~Fine, Y.~Singer, and N.~Tishby, ``\BIBforeignlanguage{en}{The {{Hierarchical
  Hidden Markov Model}}: {{Analysis}} and {{Applications}}},''
  \emph{\BIBforeignlanguage{en}{Machine Learning}}, vol.~32, no.~1, pp. 41--62,
  Jul. 1998.

\bibitem{hadar_highorder_2009}
U.~Hadar and H.~Messer, ``High-order {{Hidden Markov Models}} - {{Estimation}}
  and {{Implementation}},'' in \emph{2009 {{IEEE}}/{{SP}} 15th {{Workshop}} on
  {{Statistical Signal Processing}}}, Aug. 2009, pp. 249--252.

\bibitem{chen_chord_2012}
R.~Chen, W.~Shen, A.~Srinivasamurthy, and P.~Chordia, ``Chord {{Recognition
  Using Duration}}-{{Explicit Hidden Markov Models}},'' in \emph{13th
  {{International Society}} for {{Music Information Retrieval Conference}}
  ({{ISMIR}})}, Porto, Portugal, Oct. 2012.

\bibitem{rabiner_tutorial_1989}
L.~R. Rabiner, ``A {{Tutorial}} on {{Hidden Markov Models}} and {{Selected
  Applications}} in {{Speech Recognition}},'' \emph{Proceedings of the IEEE},
  vol.~77, no.~2, pp. 257--286, 1989.

\bibitem{goto_rwc_2002}
M.~Goto, H.~Hashiguchi, T.~Nishimura, and R.~Oka, ``{{RWC Music Database}}:
  {{Popular}}, {{Classical}} and {{Jazz Music Databases}}.'' in \emph{3rd
  {{International Conference}} on {{Music Information Retrieval}} ({{ISMIR}})},
  Paris, France, 2002.

\bibitem{digiorgi_automatic_2013}
B.~Di~Giorgi, M.~Zanoni, A.~Sarti, and S.~Tubaro,
  ``\BIBforeignlanguage{English}{Automatic chord recognition based on the
  probabilistic modeling of diatonic modal harmony},'' in
  \emph{\BIBforeignlanguage{English}{Proceedings of the 8th {{International
  Workshop}} on {{Multidimensional Systems}}}}, Erlangen, Germany, Sep. 2013.

\bibitem{burgoyne_expert_2011}
J.~A. Burgoyne, J.~Wild, and I.~Fujinaga, ``An {{Expert Ground Truth Set}} for
  {{Audio Chord Recognition}} and {{Music Analysis}}.'' in \emph{12th
  {{International Society}} for {{Music Information Retrieval Conference}}
  ({{ISMIR}})}, Miami, USA, Oct. 2011.

\bibitem{korzeniowski_largescale_2018}
F.~Korzeniowski, D.~R.~W. Sears, and G.~Widmer, ``A {{Large}}-{{Scale Study}}
  of {{Language Models}} for {{Chord Prediction}},'' in \emph{2018 {{IEEE
  International Conference}} on {{Acoustics}}, {{Speech}} and {{Signal
  Processing}} ({{ICASSP}})}, Calgary, Canada, Apr. 2018.

\end{thebibliography}
